\documentclass[runningheads]{llncs}

\usepackage{amssymb}
\usepackage{graphicx}

\usepackage{url}
\urldef{\mailsa}\path| nitinvatsa@gmail.com|

\begin{document}

\mainmatter  

\title{NNRU, a noncommutative analogue of NTRU}


%
%
\author{Nitin Vats}
%


\institute{
Indian Institute of Science, Bangalore, India\\
\mailsa\\
\url{}}

%
%

\toctitle{Lecture Notes in Computer Science} \tocauthor{Authors'
Instructions} \maketitle

\begin{abstract}
 NTRU  public key cryptosystem is well studied  lattice-based
Cryptosystem along with Ajtai-Dwork and GGH systems. Underlying
 NTRU is a hard mathematical problem of finding short vectors in a
certain lattice. (Shamir 1997) presented a lattice-based attack by
which he could find the original secret key  or alternate key.
Shamir concluded if one designs a variant of NTRU where the
calculations involved during encryption and decryption are
non-commutative then the system will be secure against Lattice
based attack.This paper presents a new cryptosystem with above
property and we have proved that it is completely secure against
Lattice based attack. It operates in the non-commutative ring
$\textbf{M}=M_k(\mathbb{Z})[X]/(X^n - I_{k\times k})$, where $
\textbf{M}$ is a matrix ring of $k\times k$ matrices of
polynomials in \(R=\mathbb{Z}[X]/(X^n-1)\). Moreover We have got
speed improvement by a factor of \(O(k^{1.624})\) over NTRU for
the same bit of
information.\\


\end{abstract}
\section{Introduction}
The first version of NTRU was proposed by (Hoffestein 1996). It
has been assessed recently as the fastest public key
cryptosystem~\cite{narb}. Its strong points are short key size,
and speed of encryption and decryption. Two assets of crucial
importance in embarked application like hand held device and
wireless systems . The description of NTRU system is given
entirely in terms of quotient ring of Integer polynomials. The
most expected attack on this system is Lattice-based attack. The
NTRU public key cryptosystem \cite{narb} relies for its security
on the presumed difficulty of solving the shortest\cite{tflc,drmf}
 and closest vector
problem in certain lattices related to the cyclotomic ring
\(\mathbb{Z}[X]/(X^n-1)\). Lattices have been studied by
cryptographers for quite some time,both in the field of
cryptanalysis and as a source of hard problems on which to build
encryption schemes\cite{narb}.\\

By lattice attack our aim is to find the original key or an
alternative key which can be used in place of original key to
decrypt ciphertext with some more computational
complexity\cite{lant}.  We construct a lattice whose elements will
corresponding to alternative key. If we get a vector as short as
original key, we can easily decrypt but even if we find a vector
that is two or three times bigger,  we can partially decrypt it by
adding the pieces to get the whole. So added security can be
achieved by increasing the dimensions of the lattice but it will
decrease the speed for encryption and decryption that is the key
property of NTRU.

In this paper we present another variant of NTRU, we will call it
NNRU. Our focus involves extension to noncommutative groups
instead of using group algebra over \(\mathbb{Z}_n\)(that is, the
ring \(\mathbb{Z}_q[X]/(X^n-1)\) ).

NNRU operates in the ring of $k$ by $k$ matrices of $k ^2$
different polynomials in \(R=\mathbb{Z}[X]/(X^n-1)\) ). As matrix
multiplication in NNRU is strictly nonabelian. Adversary will have
to find out two ring elements. So search space will be square
times than that of NTRU. In section 5 we have shown that NNRU is
completely secure against lattice attack that was more likely on
NTRU and its varients.We can compare an instance of NTRU by
putting $n(k^2)=N$ . Encryption and decryption in NTRU needs
\(O(N^2)\) or \(O(n{k^4})\) operations for a message block on
length of $N$ but in NNRU for same bit of information we need
 \(O(n{k^{2.376}})\) operations if we use coppersmith algorithms
 for matrix multiplication. that is considerable
speed improvement over original NTRU. Inversion of polynomial
matrix can be done quickly with less memory-expense by the
algorithm suggested in \cite{inver}. Moreover polynomial matrix
computations can be solved in \(\tilde{O}(n{k^e})\)by reducing
polynomial matrix multiplication to determinant computation and
conversely, under the straight line model \cite{mult}. Here
\(\tilde{O}\) denotes some missing $log(nk)$ factors and $e$ is
exponent of matrix multiplication over $R$.

The paper is organized as follows. Section 2 gives some notation
and norm estimation, that help our analysis . In section 3 we
briefly sketch NNRU cryptographic system. In section 4 we discuss
constraints for parameters. Details of the security analysis of
NNRU system is given in sections 5. Section 6 shows performance
analysis and comparison with NTRU.

\vspace{2mm}

\section{Notations}
 All computations in NNRU are performed in the ring $\textbf{M}
= M_k(\mathbb{Z})[X]/(X^n - I_{k\times k})$, where $ \textbf{M}$
is a matrix ring of $k\times k$ matrices of elements in the ring
\(R=\mathbb{Z}[X]/(X^n-1\)). An element
\(a_0+a_1x+...+a_{n-1}x^{n-1}\) of \(R\) can be represented as
$n$-tuple of integers \([a_0,a_1,...,a_{n-1}]\). Addition in \(R\)
is performed componentwise, and multiplication is a circular
convolution.

\subsection{Norm Estimation}

We define width of an element $M \in$ \textbf{M} to be
\[
\|M\|_\infty = \mbox{Max(coeff.in polys.}m \in M) -
\mbox{Min(coeff.in polys.}m \in M)\] The width of matrices $M \in$
\textbf{M} is difference between maximum and minimum coefficient
in any of $k^2$ polynomials of it.
 We say a matrix $M \in \textbf{M}$ is short if\\
$$\|M\|_\infty \leq p. $$
The width of the product of two matrices is also be short as it is
very less than $q$, though it may be slightly more than $p$. We
define width of the polynomial $r \in R$ to be

\[
\|r\|_\infty =\mbox{Max(coeff. in}\hspace{2mm} r) -
\mbox{Max(coeff. in}\hspace{2mm} r)\]

Similarly the polynomial $r$ is said to be short if $$\|r\|_\infty
\leq p.$$

Basically width of $M$ or $r$ is a sort of $L^{\infty}$ norm on
$\textbf{M}$ or $R$ respectively. In this paper we are essentially
using all calculation on the $L^2$ norm to produce an estimate of
its $L^{\infty}$ norm. For precisely evaluating the properties we
need to estimate $L^{\infty}$ but $L^2$ norm is comparatively easy
to estimate. We are giving a proposition between $L^{\infty}$ and
$L^2$ norm by which we can do all calculations on $L^2$ norm and
estimate on $L^{\infty}$ norm. It is based on experiments and
suggestions due to Don Coppersmith\cite{narb}\\

Let $\|r\|$ be the $L^2$ norm for a random polynomials $r$. Then
following proposition is true for random polynomials $r_1, r_2 \in
R$ with small coefficients .
\begin{eqnarray}
&\|r_1*r_2\|& \approx \|r_1\|.\|r_2\|\nonumber \\
{\mbox{ and}} \;\;\;&\|r_1*r_2\|&_\infty \approx
\gamma\|r_1\|.\|r_2\|\hspace{2mm}\mbox{where}, \gamma <
0.15\hspace{2mm} \mbox{ for}\hspace{2mm} n < 1000.
\end{eqnarray}
 Now we define a centered $L^2$ norm on $\textbf{M}$.We denote it
 by the notation $\|M\|$.\\

 \begin{eqnarray*}
\|M\| = \sqrt{{\sum_{({\rm polys.}m \in M)}\sum (\mbox{Coeff. in}
\hspace{1mm}} m-\mu)^2)}\hspace{2mm}
\end{eqnarray*}
\vspace{2mm}

$\mbox{where}\hspace{4mm}\mu=\frac{1}{nk^2}\left(\sum_{({\rm
polys.}m \in M)}\sum (\mbox{Coeff. in \hspace{1mm}} m)\right)$
  is the average of all coefficient in all the
polynomial in matrices $M$. Its value will be close or equal to
zero. Equivalently $ \|M\|/\sqrt{nk^2}$ is standard deviation of
the coefficients of the polynomials in $M \in \textbf{M}$. In this
paper we do analysis on $L^2$ centered norm of $M$ and can deduce
results on $L^{\infty}$ norm by using result (1).\\

The proposition (1) can be extended to the centered $L^2$ norm on
$\textbf{M}$. Consider any $\kappa > 0$ there are constants
$\gamma_1,\gamma_2 > 0 $ and two matrices $M_1,M_2\in \textbf{M} $
We therefore express
\begin{eqnarray}
\|M_1*M_2\| &\approx& \|M_1\|.\|M_2\|\nonumber \\
\mbox{and}\hspace{2mm}\gamma_1\|M_1\|.\|M_2\|&\leq&
\|M_1*M_2\|_\infty \leq \gamma_2\|M_1\|.\|M_2\|
\end{eqnarray}

On the basis of experimental evidence and due to  Don
Coppersmith\cite{narb}, The preposition holds good with
probability greater than $1-\kappa$ for small $\kappa$. It can be
shown experimentally that even for larger value of $nk^2$, the
value of $\gamma_1/\gamma_2$ is somewhat between zero and one
(moderately larger than zero).
\subsection{Sample Spaces}

NNRU cryptosystem depends on four positive integer parameters
$(n,k,p,q)$ with $p$ and $q$ relatively prime and four sets of
matrices $(L_f, L_c, L_\phi, L_m)\subset \textbf{M}$. Note that
$q$ will always be considered much larger than $p$. In this paper,
for ease of explanation, we stick to $p=2$ or 3, and $q$ ranges
between $2^8$ to $2^{11}$. When we do Matrix multiplication modulo
$p$ (or $q$), we mean to reduce the
coefficients of the polynomial in matrices modulo $p$ (or $q$).\\

The set of matrices ($L_f, L_c, L_\phi,L_m $) consists of all
matrices of polynomials in the ring \(R=\mathbb{Z}[X]/(X^n-1\)).
The set of matrices ($L_f, L_c, L_\phi $) contains polynomials
from the set of polynomials $L(d_1,d_2)$
 \begin{eqnarray*}
L(d_1,d_2) \stackrel{\mathrm{def}}{=} \{u \in R &| u\hspace{2mm}
\mbox{has}\hspace{2mm} d_1 \mbox{ coeff. equal}\hspace{2mm} 1, d_2
\hspace{2mm}\mbox{coeff. equal} -1,
 &\mbox{and}\hspace{2mm} \mbox{rest}\hspace{2mm} 0\}.
\end{eqnarray*}

 where, \(d_1 =d_2 < n/2\) ~~~~~or~~~~~~~ \(d_1 = d_2  \approx n/p
 \)

 The space of message $L_m$ consists of all matrices of
 polynomials with coefficients modulo $p$.We therefore express\\

 $L_m \stackrel{\mathrm{def}}= \{M \in \textbf{M} |$polynomial in $M $ has
 coeff. lying between $-\frac{p-1}{2}$
 and $ \frac{p-1}{2}$\}.\\

 Here we explain individually the meaning and compositions of the
 all four sets of matrices $(L_f, L_c, L_\phi, L_m)\subset
 \textbf{M}$:
\begin{enumerate}
\item$L_f$ with elements $f$ and $g$, and $L_\phi$ with elements
$\phi$ consist of small matrices of polynomials $f$ and $g$, are
used to compose private key while $\phi$ will be used as blinding
value for each encryption. $L_f$  must satisfy the requirement to
have inverse modulo $p$ and modulo $q$.

\item element $w$ and $c$ belongs to matrix set $L_w$ and $L_c$
respectively. $L_c$ should satisfy the requirement that to have
inverse modulo $p$ . $w$ and $c$ are used to construct public key.

\item  the set of message $L_m$ consist of matrices of polynomials
with coefficients modulo $p$ .We therefore express
\begin{eqnarray*}
L_m= \left\{M \in \textbf{M}\left|\right ( \mbox{Polys.})
\hspace{2mm} (\mbox{in})\hspace{2mm} M \in \left(-\frac{p-1}{2}
\cdots \cdots\frac{p-1}{2}\right)^n \subseteq R\right\}
\end{eqnarray*}

\end{enumerate}

\section{The NNRU System}
\subsection{Key Creation}
To create a NNRU public/private key pair Bob randomly chooses $f,
g \in L_f$ and $w \in L_w$ and $c \in L_c$. Matrices $f$ must
satisfy additional requirement to have inverse modulo $p$ and $q$.
Matrices $g$ and $c$ should have inverse modulo $p$ . We denote
these inverses by notation $F_p$, $F_Q$, $G_P$, $C_p$ respectively.\\

 \centerline{\(f\ F_q \equiv I(\bmod q)\)\hspace{3mm} {\rm and}
\hspace{3mm}\(g \ G_p \equiv I(\bmod p)\)} \vspace{.2cm}
\centerline{\(G_q\ g \equiv I({\rm mod}q)\)\hspace{3mm} {\rm and}
\hspace{3mm}\(C_p\ c \equiv I(\bmod p)\)}

 Bob next computes the matrices

\begin{eqnarray}
h &\equiv& w G_q \hspace{1mm}(\bmod q)\\
H &\equiv& F_q c \hspace{1mm}(\bmod q)
\end{eqnarray}

Bob publish the pair of matrices $(h,H)\in \textbf{M} $ as his
public key, retaining ($f, g, c$) as his private key. Polynomial
$C_p$ and $G_p$ is simply stored for later use.

\subsection{Encryption}

Suppose Alice(the encryptor)wants to send a message to Bob (the
decryptor). Alice selects a message $m$ from the set of plaintext
$L_m$. Next, Alice randomly choose a matrices $\phi \in L_\phi $
and use, Bob's public key $(h, H)$ to compute (the ciphertext $e$)

 \[e\equiv p\phi h + H m \hspace{1mm} (\bmod q)\]

Alice then transmit $e$ to Bob. A different random choices of
blinding value $\phi$ is made for each plaintext $m$ .

\subsection{Decryption} To decrypt the cipher text, Bob first compute
\begin{eqnarray*}
A&\equiv& feg\hspace{1mm}(\bmod q)\nonumber\\
A&\equiv& f(p\phi h +H m )g\hspace{1mm}(\bmod q)\nonumber\\
A&\equiv& f p\phi h g +f H m g\hspace{1mm}(\bmod q)\nonumber\\
 A&\equiv& p f \phi w G_q g +f F_q c m g\hspace{1mm}(\bmod q)\nonumber\\
 A&\equiv& p f \phi w + c m g\hspace{1mm}(\bmod q)\nonumber
\end{eqnarray*}

Where he choose the coefficients of the polynomials of the
matrices $A$ to lie in interval of $-q/2$ to $q/2$ . Why
decryption works? Matrices $\phi$, $g$, $f$, $m$, $c$ and $w$ have
polynomials with small coefficients and $p$ is much smaller than
$q$. It is highly probable for the appropriate parameter choice of
the members, matrices $p f \phi w + c m g $, before reducing
mod\hspace{1mm}$q$, has polynomials with coefficients of absolute
value less than $q/2$.
 Bob next computes the matrices $B$
\begin{eqnarray}
 B&\equiv& A (\bmod p)\nonumber\\
 B&\equiv& cmg (\bmod p)\nonumber
\end{eqnarray}
He reduces each coefficient of the element of $A$ to modulo $p$ .
Finally Bob uses his other private keys $C_p$ and $G_p$ to recover
the original message.
\begin{eqnarray}
C&\equiv&C_p cmg G_p(\bmod p)\nonumber\\
 C&\equiv& m(\bmod p)\nonumber
\end{eqnarray}

The matrix $C$ will be the original message $m$ as
\[\mbox{polynomial in } m \in {\left(-\frac{p-1}{2} \cdots
\cdots\frac{p-1}{2}\right)}^n \subseteq R\]

\section{Parameter Constraint}
Our selection is based on the following three requirements
\begin{enumerate}
\item $f\phi w$ and $cmg$ should be small in order for decryption
to work. \item Appropriate selection of $f$, $g$ and $c$ prevent a
private key attack. \item Appropriate selection of $\phi$ and $m$
prevent plain text attack.
\end{enumerate}

The key point is that decryption will only work if $f\phi w$ and
$cmg$ are not too large so we want to keep $|p f\phi w +
cmg|_\infty$ should be small. For security reasons, it is
important that $w$, remains secret from attacker. On average $|w|
\approx |m|$. this type of selection follows
\[|p f \phi w| \approx |cmg|\]

As already described that we are selecting $f$, $g$ from $L_f$,
$c$ from $L_c$ and $w$ from $L_w$ , $m$ from $L_m$ which gives
$d_1= d_2 \approx n/p$ ; that ensure to maximize the number of
possible choices for polynomials of these matrices.

\section{Cryptanalysis}

\subsection{Brute Force Attacks} To decrypt the cipher text,
attackers need to know the private key $f$, $g$ and $c$ correctly.
Attacker can try all possible $f, g \in L_f$ so that $h
g\hspace{1mm} (\bmod q)$ should have polynomials with small
entries or by finding all $g \in L_f$ and testing if $f
H\hspace{1mm}(\bmod q)$ have polynomial with small entries. Out of
these small $f H\hspace{1mm}(\bmod q)$, one will be
$c\hspace{1mm}(\bmod q)$. So attacker need to search pair of $(f,
g)$. $f$ and $g$ are determined by $2k^2$ polynomials, each of
them having maximum degree $(n-1)$. so the number of possible $(f,
g)$ pairs are
\[\mbox{Key Security}={\left[\frac{n!}{{(n- 2d_f)!{d_f!}^2}}\right]}^{2k^2}\]

 Here $d_f$ and $d_\phi$ are defined by assuming $L_f$ and $L_\phi$ contains polynomials from the set of
 polynomials $L(d_f, d_f)$ and $L(d_\phi, d_\phi)$ respectively.
 By analogy, the same attack can also be done against a given
 message by testing all possible $ \phi \in L_ \phi $ and search
 for the matrices $e-\phi h (\bmod Q)$ which contains polynomials with small
 entries. So individual message security is defined by
\[\mbox{Message Security}={\left[\frac{n!}{{(n- 2d_\phi )!{d_\phi!}^2}}\right]}^{2k^2}\]

A meet-in-middle attack was proposed by Andrew Odlyzko \cite{idf}
for NTRU and developed by Silverman. This attack can also be used
against NNRU. The attack need a lot of storage capacity and cut
the search time by the square root.

\subsection {Multiple Transmission Attack}
This attack works if Alice sends a single message $m$ several time
using same public key but different blinding values $\phi$'s, then
the attacker eve can get the maximum bits of the message.

suppose Alice transmit the massage
\begin{eqnarray*}
e_i&\equiv&\phi_i h + H m(\bmod q) \nonumber
 \end{eqnarray*}
for $i=1,2 \dots \dots \dots \dots r$

eve can compute $(e_i-e_1)*h^{-1}(\bmod q)$. therefore recovering
$\phi_i-\phi_1(\bmod q)$. If $r$ is of moderate size (say 5 or 6),
eve will recover enough bits of $\phi_1$to apply brute force to
the rest of the bits. As polynomial of $\phi$ have small
coefficients so eve will recover exactly $\phi_i-\phi_1$, and in
the way eve will recover many of coefficients of polynomial of
$\phi_1$

due to this attack we suggest not to use multiple transmission
with further scrambling of particular (underlying) message.
However this attack will work for a single message(tha has been
multiple transmitted)not for any subsequent message.

\subsection{Lattice Attack}

The Decryptor computes
\begin{eqnarray*}
A&=& f e g \equiv p f \phi w +c m g \hspace{2mm}(\bmod q)
\nonumber
 \end{eqnarray*}

 parameter are chosen so that both $p f \phi w$ and $c m g$ are small
 enough to guarantee the entries of non modular expression
 \begin{eqnarray*}
B = p f \phi w +c m g \hspace{2mm} (\bmod q) \nonumber
 \end{eqnarray*}

 lies between $-q/2$ and $q/2$ most of the time. In this case
 decryptor can switch to compute modulo $p$ from computing modulo
 $q$ and can calculate message.
  \begin{eqnarray*}
m \equiv C_p B G_p \hspace{2mm}(\bmod p) \nonumber
 \end{eqnarray*}
we can estimate bounds on the elements of $B$ provided correct
decryption. Decryption will work only when $B$ is equal to $p f
\phi w +c m g $, not mere congruent to modulo $q$. Using
result(2)we can say the following

\[\parallel p f \phi w\parallel \approx p\parallel f\parallel\parallel \phi\parallel\parallel w\parallel\]
\[\parallel c m g\parallel \approx\parallel c\parallel\parallel m \parallel\parallel g\parallel\]
Assuming vectors $p f \phi w$ and $c m g $ to be nearly
orthogonal, we can write

 \begin{eqnarray}
\parallel B\parallel^2 \approx p^2\parallel
f\parallel^2\parallel \phi\parallel^2\parallel w\parallel^2\ +
\parallel c
 \parallel^2\parallel m \parallel^2\parallel g\parallel^2
\end{eqnarray}

 decoding will fail if any coefficient of polynomial of $B$ will
 more than $q/2$ in absolute value. Make the second assumption that
 the entries of polynomials in matrices $B$ are normally
 distributed with mean zero and standard deviation
 $\sigma \approx \frac{\parallel B\parallel}{\sqrt{nk^2}} $.
 Analogues to shamir's results for NTRU \cite{narb}, Experiments
 suggests the fact that the probability of correct decoding is
 high for small ratio of $\sigma $ to $q/2$. We can say that reliability of decoding is directly
proportional to the ratio of $\sigma \approx \frac{\parallel
B\parallel}{\sqrt{nk^2}} $ to $q$\\

 Equation (5) gives an estimate of the value of $B$ in terms
of $f,w,c$ and $g$. Let us consider the case in which attacker
 can use an alternate matrices $f'$ in place of original $f$ and $g'$ in place of $g$.
  Upon calculate from a value of $w'$ from equation (3)  and $c'$ from equation
  (4), an estimate of $\parallel B'\parallel$ can be calculated by equation (5). If
this $\parallel B'\parallel$ is comparable to $\parallel
B\parallel$, then it is not tough to recover message using $f'$
and $g'$ so consider

\[\parallel B\parallel^2 \approx p^2\parallel
f\parallel^2\parallel \phi\parallel^2\parallel w\parallel^2\ +
\parallel c
 \parallel^2\parallel m \parallel^2\parallel g\parallel^2\]

 Assume $\parallel \phi\parallel$ and $\parallel m
 \parallel$ to be held constant at a typical value, and putting
 $\lambda= \parallel m \parallel /{p \parallel \phi \parallel}$,
 putting the value of $\lambda$ in above equation, we therefore
 left with

 \begin{eqnarray*}
 {\sigma}^2 = \frac{{\parallel B'\parallel}^2} {nk^2} \approx \left(\frac{p^2{\parallel
\phi \parallel}^2}{nk^2}\right)({\parallel f'
\parallel}^2{\parallel w'\parallel}^2 + {\lambda }^2 {\parallel c' \parallel}^2{\parallel g'
\parallel}^2  )\nonumber\\
\end{eqnarray*}

We can attack this cryptosystem if we can make a lattice $L$ in
which squared norm of an element being
\[\parallel f \parallel^2\parallel w \parallel^2+ \parallel c
 \parallel^2\parallel g \parallel^2\]

  In other words if we can construct a lattice from public key pair $h, H$
  in which vector ($fw, cg$) lies or if we show vectors $fw$ and
  $cg$ to be same linear transformation of public key vectors. In following analysis we show that
   we can't make such lattice that will generated by public key and contain vectors ($fw, cg$).

 Encrypted message is left multiplied by $f$ and right multiplied
by $g$. $f w$ and $c g $ are produced by following transformation
on public keys.\\
\[T_{f,g}(1): 1 \mapsto f\ g\]
We can define $T_{f, g}: M \rightarrow M$ be the linear map
 \begin{eqnarray}
 &h&\mapsto f h g \hspace {3mm} \mbox{or} \hspace {3mm}h \mapsto f w \\
  &H&\mapsto f H g \hspace {3mm} \mbox{or} \hspace {3mm}H \mapsto c g
  \end{eqnarray}

 For further analysis Let us consider the definition of a lattice. Let ${\bbbr}^m$ be
 the $m$-dimensional Euclidian space. A lattice in ${\bbbr}^m$is the
 set
 \begin{eqnarray*}
\textit{L}(b_1,b_2,b_3, \dots \dots ,b_n)=\left\{\sum_{i=1}^{n}x_i
b_i : x_i \in \bbbz \right\}\nonumber
\end{eqnarray*}
 of all integer combination of $n$-linear independent vectors
 $\left\{b_1,b_2,b_3, \dots \dots ,b_n\right\}$ in ${\bbbr}^m(m \geq
 n)$.
 Here we try to make a Lattice of dimensions $2nk^2 \times 2nk^2$
 with basis vectors produced by the cyclic shift of the
 coefficients of polynomial of the matrices $h$ and $H$.
 Attacker can crack the system provided the Lattice
 contains vector ($fw, cg$).

One can conclude by linear transformation shown in equation (6)
and (7) that the lattice attack is possible if and only if one can
make a lattice with public key vectors ($h,H$) which contains
vector ($fw,cg$) or if following transformation is linear
\begin{eqnarray}
 (h,H)\mapsto (f w,c g) \hspace {3mm}
\end{eqnarray}

In following analysis we show transformation $ h \mapsto fhg$ is
not linear. Similarly it follows $ H \mapsto fhg$ and
$(h,H)\mapsto (f w,c g) $ can not be linear.

Consider the multiplication of the matrices $f.h.g=fw$, where each
matrix ($f,g,h,fw$)having $k^2$ short polynomials as elements
 \vspace{4mm}
\begin{eqnarray*}
\left[
  \begin{array}{ccc}
    f_{1} & \cdots & f_{k} \\
    \vdots & \ddots & \vdots \\
    f_{k(k-1)} & \cdots& f_{k^2}
  \end{array}
  \right]\left[
  \begin{array}{ccc}
    h_1 & \cdots & h_k \\
    \vdots & \vdots & \vdots \\
    h_{k(k-1)} & \cdots& h_{k^2}
  \end{array}
\right]\left[
  \begin{array}{ccc}
    g_1 & \cdots & g_k \\
    \vdots & \vdots & \vdots \\
    g_{k(k-1)} & \cdots& g_{k^2}
  \end{array}
\right]=\left[
  \begin{array}{ccc}
    fw_{1,1} & \cdots & fw_{1,k} \\
    \vdots & \vdots & \vdots \\
    fw_{k,1} & \cdots& fw_{k,k}
  \end{array}
\right]
\end{eqnarray*}

 \begin{eqnarray*}
 (fw)_{1,1} = &g_1&f_1h_1 + g_{k+1}f_1h_2 + g_{2k+1}f_1h_3 + \cdots +
g_{k(k-1)+1}f_1h_k + g_1f_2h_{k+1}\nonumber\\
&+&\cdots +g_{k(k-1)+1}f_2h_{2k}+ \cdots +
 g_{k(k-1)+1}f_kh_{k^2}\nonumber\nonumber\\
(fw)_{1,2} = &g_2&f_1h_1 + \cdots \cdots  \cdots \cdots \cdots
\cdots \cdots \cdots
 \cdots \cdots \cdots  \cdots \cdots +
 g_{k(k-1)+2}f_kh_{k^2}\nonumber\\
&\vdots& \\
(fw)_{k,k} = &g_k&h_1f_{k(k-1)+1} + g_{2k}h_2f_{k(k-1)+1} + \cdots
+ g_{k^2}h_kf_{k(k-1)+1} + \cdots g_{k^2}h_{k^2}f_{k^2}\nonumber
\end{eqnarray*}

So general term can be represented as
\[(fw)_{i,j} =
\sum_{l=k(i-1)+1}^{ki}\sum_{s=0}^{k-1}f_l(g_{j+sk})\left(h_{(1+s)(l-k(i-1))}\right)\]
or, we can represent \((fw)_{i,j} = \sum f_ug_vh_z\ = \sum
U_zh_z\) where, $u$, $v$, and $z$ are according to the
relationship shown above,
\begin{eqnarray*}
\mbox{Here}\hspace{2mm}i,j \in [1 \;\; k^2];\;\;u, v \in [1 \;\;
k^2];\;\;z \in [1 \;\; k^4]\nonumber
\end{eqnarray*}

As all $U_z$ are different so we can not find a row vector
$S_i=(s_1,s_2,\dots \dots, s_{k^2})$ that will produce vector $fw$
on multiplying with a Lattice represented by the cyclic shift of
the coefficients of polynomial of $h$. In other words if column
vectors ${v_1,v_2,\dots \dots v_{nk^2} }$ are the basis of lattice
$L(v_1,v_2,\dots \dots v_{nk^2})$, then we will have to multiply
different vector $S_i$ to each column vector $v_i$ to get $fw$. We
therefore conclude
\begin{eqnarray*}
fw \neq S_i\ L(v_1,v_2,\dots \dots v_{nk^2} ) \nonumber
\end{eqnarray*}

 Thus we proved that one cannot make a lattice by $h$
  and $H$, which contains the vectors ($fw, cg$).
 So lattice attack will not work for this cryptosystem unlike
 NTRU\cite{narb} and its variants \cite{matru}.

\section{Comparison of Security and Speed of NNRU with Other Variants of NTRU}
 Many variants of NTRU have been
    introduced till date. We present NNRU as the only variant of NTRU which operates in non-commutative ring.
    It is completely secure against Lattice attack. Moreover it gives speed improvement over
    NTRU. Brief of other variants are as follows.\\

1. \textbf{Variant with non-invertible polynomial }\cite{var}: It
operates in ring  \(\mathbb{Z}[X]/(X^N-1)\) . Size of public key
and encryption time is roughly doubled than NTRU. It is likely to
be more robust
against Lattice attack but not proved.\\

2. \textbf{MaTRU }\cite{matru}: It operates in a ring of $k\times
k$ matrices of polynomials in \(R=\mathbb{Z}[X]/(X^n-1)\) but
decryption is not non-commutative. Speed improvement is achieved
by a factor of \(O(k)\). It gives no added security against
lattice or other attacks in comparison with
NTRU. \\

 3. \textbf{CTRU }\cite{ctru}: It is analogue of NTRU, the ring of integers
replaced by the ring of polynomials \(\bbbf_2[T] \). It has been
completely cracked by linear algebra attack.\\

As \cite{var} is slow and \cite{ctru} is completely cracked so it
is obvious to give more attention to the study of security aspect
of MaTRU. Here we present meet-in-middle attack on MaTRU and show
that the MaTRU system is not more robust against this attack
compare to NTRU. This attack can't be operated on NNRU because
calculations involved in decryption are non-commutating.
\cite{meet} shows meet-in-middle attack on NTRU. We show that
similar attack can be applied on MaTRU. \\
Applying same notations as in \cite{matru} let us consider Second
block of MaTRU Lattice \cite{matru}.

\[w~ (\bmod q) = \left[\begin{array}{c}
\gamma_{0,0}\\
\gamma_{0,1}\\
\vdots\\
\vdots\\
\gamma_{k-1, k-1}
\end{array}\right]^T
\left[
  \begin{array}{c}
    \framebox[1in][c]{$h$}\vspace{0.1cm}\\
    \framebox[1in][c]{$h \gg 1$}\vspace{0.1cm}\\
    \framebox[1in][c]{$h \gg 2$}\vspace{0.1cm}\\
    \vdots\\
    \framebox[1in][c]{$h \gg k^2-1$}
  \end{array}\right]\]
$nk^2$ coefficients of $w$ can be achieved by multiplying row
vector $\gamma$ to matrix $h$. Idea is to search for $\gamma$ in
the form $\gamma_1 || \gamma_2$, where $\gamma_1 $ and $ \gamma_2$
are each of ${nk^2}/2$ length with $d/2$ ones and ``$||$'' denotes
concatenation, and then to match (${\gamma_1}* h$) against
($-{\gamma_2}* h$), looking for $(\gamma_1,\gamma_2)$ so that the
corresponding coefficients have approximately the same value. The
above relationship can be written as
\begin{eqnarray*}
\Rightarrow(\gamma_1*h)_i &=& \{0,1\} - (\gamma_2*h)_i\;(\bmod
q)\forall_i \nonumber
\end{eqnarray*}
where, the $a_i$ notation denotes the $i^{th}$ entry in $a$.\\
This equation is similar to what we get for NTRU \cite{meet}.
\begin{eqnarray*}
\Rightarrow(f_1*h)_i &=& \{0,1\} - (f_2*h)_i\;(\bmod q)\forall_i
\nonumber
\end{eqnarray*}
We can operate the attack same as \cite{meet}. Assuming $nk^2=N$
and $d$ are number of ones in $\gamma$. Similar to \cite{meet},
One can easily find that the expected running time and storage
space required for this method (this value is equal to what we get
for NTRU)is ${\left(\begin{array}{c}
N/2\\d/2\end{array}\right)}/{\sqrt{N}}$. Further one can also
apply meet-in-middle attack on MaTRU followed
 by Linear algebra attack. Lattice in \cite{matru} can also be represented as modular
 equation \( \gamma(y)*h(y) \equiv w(\bmod q)(\bmod
(y^{k^2}-1))\). It can also be written as
\[\gamma(y)*h(y)= w + qu\]
where, \(u = u_{0,0} + u_{0,1} + \cdots + u_{k-1,k-1}y^{k^2-1}\)
and, \(u_{i,j} \in \mathbb{Z}[X]/(X^n-1)\).
Above system of linear equations consist of $3nk^2-1$ variable in
$nk^2-1$ linear equations. If $nk^2-1$ is not fairly large than
the system of linear equations can be used to reduce an exhaustive
search to a space of size  $2^{nk^2-1}$. further one can set up a
meet in middle search to reduce the running time to
$\emph{O}$$(2^{(nk^2-1)/2})$.
\section{Performance Analysis and Comparison with NTRU}

Here we present the theoretical operating specification of NNRU
and compare the complexity of different operation with standard
NTRU PKCS. NNRU cryptosystem depends on four positive integer
parameters $(n,k,p,q)$ with $p$ and $q$ relatively prime and four
sets of matrices $(L_f, L_c, L_\phi, L_m)\subset \textbf{M}$ .The
properties of NTRU \cite{narb} is defined in terms of parameters
$(N,p,q)$. We compare two systems for the same size of plaintext
blocks by setting $N=nk^{2}$.

 \begin{tabular}{|c|c |c|c|c|c|c|c|c|}

\hline Characteristics\hspace{2mm} & NTRU  & NNRU \\
\hline
 Plain text Block & $N\log_{2}p$ bits& $nk^{2}\log_{2}p$ bits \\ 
 Encrypted Text Block & $N\log_{2}q$ bits& $nk^{2}\log_{2}q$ bits  \\ 
 Encryption Speed & $O({N^2})$ operations& $O({n^{2}k^{3}})$ operations \\ 
Message Expansion &$\log_{p}q$~to~1&  $\log_{p}q$~to~1\\
 Private Key Length &  $2N\log_{2}p$ bits& $2nk^{2}\log_{2}p$ bits \\ 
Public Key Length &  $N\log_{2}q$ bits &   $2nk^{2}\log_{2}q$ bits \\ 
 Lattice Security  &  \(2\left(\frac{\pi^2ae^2}{3Nq^2}\right)^{\frac{1}{4}}\)& Totally secure against lattice attack\\ 

\hline
\end{tabular}
\vspace{4mm}

${}^1$Since NNRU perform two-sided multiplication during
decryption process,
 so constant factor will about twice that of standard NTRU

${}^2$ For message security $d_g$ will be replaced by $d$ for NTRU
and $d_f$ to $d_\phi$ for NNRU Cryptosystem

If we compare the size of public/private key, NNRU needs two
public keys each of them is double in length that of NTRU public
key while the size of private key is same. NNRU gives significant
speed improvement over standard NTRU.
 We can compare an instance of NTRU by putting $n(k^2)=N$ .
Encryption and decryption in NTRU needs \(O(N^2)\) or
\(O(n{k^4})\) operations for a message block on length of $N$. In
NNRU the same bit of information requires \(O(n{k^{2.807}})\) or
\(O(n{k^{ 2.376}})\) operations if we use Strassen's or
coppersmith algorithms for matrix multiplication respectively. We
can further reduce the number of operations if we use FFT for
polynomial multiplication. In this case it will be as small as
\(O( {k^{2.376}n \mbox{log} {n}})\), which is considerable speed
improvement over original NTRU. It is faster than RSA which needs
\(O(N^3)\)operations for encryption and decryption.
\section{Conclusion}
Our motivation for NNRU results from various suggestions given  by
Shamir and other researchers in their papers for extensions to
non-commutative groups. We studied NTRU over ring
$\mathbb{F}_{2}(T)[X]/(X^{n}-1)$ but we found that, the variant
\cite{ctru} is secure against Popov Normal Form attack but
completely insecure against linear algebra based attacks . Here we
follow group algebra over strictly non-commutative groups. Lattice
attack is biggest threat to NTRU. It is expected that new lattice
reduction technique will be discovered over time and will be able
to reduce number of arithmetic operations involved in it. It is
natural to study an analogue of NTRU in the given context and find
the possibilities in terms of security against Lattice attack and
any improvement in terms of speed. NNRU is completely secure
against Lattice attacks with significant speed improvement.
Further research can be done in the direction of finding the
possibilities of any other type of attack or further improvement
and generalization of NNRU Cryptosystem.

%
%

%
\end{document}